%% file: stock_2017.tex
\documentclass{ws-rv961x669}
\usepackage{ws-rv-van}     % numbered citation/references (default)
\usepackage{ws-rv-thm}     % comment this line when `amsthm / theorem / ntheorem` package is used
\usepackage{subfigure}     % required only when side-by-side / subfigures are used
\makeindex
\begin{document}

\chapter[The QCD Phase Diagram from Statistical Model Analysis]{The QCD Phase Diagram from Statistical Model Analysis\label{ra_ch1}}

%\author[Francesco Becattini, Marcus Bleicher, Jan Steinheimer and Reinhard Stock]

\author{Francesco Becattini}

\address{Universita di Firenze and INFN Sezione di Firenze, Firenze}

\author{Marcus Bleicher, Jan Steinheimer}
\address{Frankfurt Institute of Advanced Studies (FIAS), Frankfurt}

\author{Reinhard Stock}

\address{Frankfurt Institute of Advanced Studies (FIAS), Frankfurt\\
Institut fuer Kernphysik, Goethe Universitaet, Frankfurt}

\begin{abstract}
Ideally, the Statistical Hadronization Model (SHM) freeze-out curve should reveal the QCD parton-hadron phase transformation line in the ($T$,$\mu_B$) plane. We discuss the effects of various final state interaction phenomena, like baryon-antibaryon annihilation, core-corona effects or QCD critical point formation, which shift or deform the SHM freezeout curve. In particular, we present a method to remove the annihilation effects by quantifying them with the microscopic hadron transport model UrQMD. We further discuss the new aspects of hadronization that could be associated with the relatively broad cross-over phase transformation as predicted by lattice-QCD theory at low $\mu_B$. That opens up the possibility that various observables of hadronization, e.g. hadron formation or susceptibilities of higher order (related to grand canonical fluctuations of conserved hadronic charges) may freeze out at different characteristic temperatures. This puts
into question the concept of a universal \textit{(pseudo-)critical} temperature, as does the very nature of a cross-over phase transformation.
\end{abstract}

%\markboth{Even Page Header}{Odd Page Header} % Customized running heads

\body

%\tableofcontents

\section{Introduction}
\label{sec:1}
The phase diagram of strongly interacting matter, as addressed by the 
thermodynamics of Quantum Chromodynamics (QCD) theory, represents one of the 
open problems of the Standard Model of elementary interactions. Thermal QCD 
confronts matter at high energy density, as was prevailing during the early 
phases of the cosmological expansion but can also be investigated (albeit not 
at macroscopic scale) in collisions of heavy nuclei at relativistic energy. One 
thus creates a hot and compressed \textit{fireball} in which, for an instant of time, 
the QCD confinement of partons into hadrons is overcome, creating
a hot partonic QCD plasma state, the so-called Quark-Gluon Plasma(QGP). It is 
the goal, both of experiment and of QCD thermodynamics, to elaborate the phase 
diagram of \textit{fireball matter}. Its most prominent feature, the transition line
between hadrons and partons, in the plane spanned by temperature T and 
baryochemical potential $\mu_B$, is located in the nonperturbative sector of QCD. 
Here, the theory can (only) be solved on the lattice\cite{1}, within the above
variables T and $\mu_B$, and has recently led to predictions of the parton-hadron
boundary line\cite{2,3}.

This line can also be addressed by experiment, in relativistic collisions of 
heavy nuclei where the primordial interaction volume is far from equilibrium at 
first; but after a small relaxation time (of order 1 fm/c, or even less) it 
settles onto the equilibrium ($T$,$\mu_B$) plane at a temperature well above the 
\textit{critical} (or pseudocritical) temperature $T_c$ that corresponds to the phase 
transformation. Expansion and cooling then take the fireball down to the QCD 
phase boundary where hadronization occurs(4), with the parameters 
($T_c$,$\mu_{B,c}$) being preserved in the relative abundances (multiplicities) of 
the various created hadronic species\cite{4,5}.  Assuming, for the moment, that 
these multiplicities are conserved throughout the final hadron-resonance 
cascade expansion, their analysis in the framework of the Statistical 
Hadronization Model(SHM)\cite{4,5,6,7,8} reveals the hadronization point, at which, 
in this ideal picture, the partons freeze-out into hadrons\cite{9,10,11}. The hadron abundances stay essentially unobliterated throughout the ensuing hadron/resonance expansion evolution. The latter fixes flow observables, spectra and Bose correlations. So there are two separate, well spaced freeze-outs, the so-called hadro-chemical and the kinetic ones.

As the primordial temperature (baryochemical potential) shifts upward 
(downward) with increasing collision energy, an ascending sequence of 
experimental energies can, thus, map a sequence of hadronization points along 
the QCD parton-hadron boundary line. At the low AGS energies, in the domain of 
$\sqrt{s} = 5$~GeV, this investigation begins at $\mu_B$ of about $550$~MeV, falling
to about $250$~MeV at top SPS energy ($\sqrt{s} = 17.3$~GeV), then on to the $20$ to 
$100$~MeV domain at RHIC energies ($\sqrt{s}$ from $60$ to $200$~GeV), and ending at 
baryochemical potential practically zero at the LHC energy of 2.76 TeV, in 
central collisions of Pb or Au projectiles\cite{6,7,12}. It is here that we meet
with the conditions prevailing in the big-bang expansion, thus recreating the 
cosmological phase transition to hadrons that occurred at a few microseconds 
time.

With these assumptions which of course need careful discussion and refinement, the subject of the present article, one of the main goals common to QCD 
theory and nuclear collision experiments --- the parton-hadron boundary
line --- comes well within reach as we shall demonstrate in the following. We 
will describe the state of the art in applying the SHM to the hadron 
multiplicity results created at the AGS, SPS, RHIC and LHC (also discussing 
difficulties arising in the application of the statistical model), and compare 
the resulting hadronization curve with the corresponding lattice QCD 
calculations.

Turning to a more detailed argumentation, we note that at present
the identification of the QCD hadronization point with the emergence of a 
hadron/resonance population that is ``born into equilibrium''\cite{9}, thus resembling 
an equilibrium canonical or grand-canonical Gibbs ensemble (the basis of the 
Statistical Hadronization Model\cite{4,5}), remains a (plausible) conjecture.
No analytic solution of the QCD Lagrangian exists in the non-perturbative 
domain. Two approaches nevertheless suggest the validity of this assumption. 
The first is a line of argument developed by Amati, Veneziano and 
Webber\cite{13,14}, in an early attempt to understand hadron production in 
electron-positron annihilation. A QCD mechanism called \textit{colour preconfinement} 
occurring toward the end of the QCD shower evolution prompts colour
neutralization into singlet \textit{clusters} or resonances, that subsequently decay
quantum mechanically onto the hadron/resonance mass spectrum. This decay is
 governed by phase space weights (Fermis Golden Rule), thus the outcome appears to be
born into equilibrium, a maximum entropy state preserving the initial quantum
numbers and the size of available phase space, represented by temperature and
baryochemical potential in the Gibbs ensemble. Indeed the hadron multiplicities
from e$^{+}$--~e$^{-}$~annihilation can be well accounted for by the SHM.
\begin{figure}[t]
%\sidecaption
\center
\includegraphics[width=.6\textwidth]{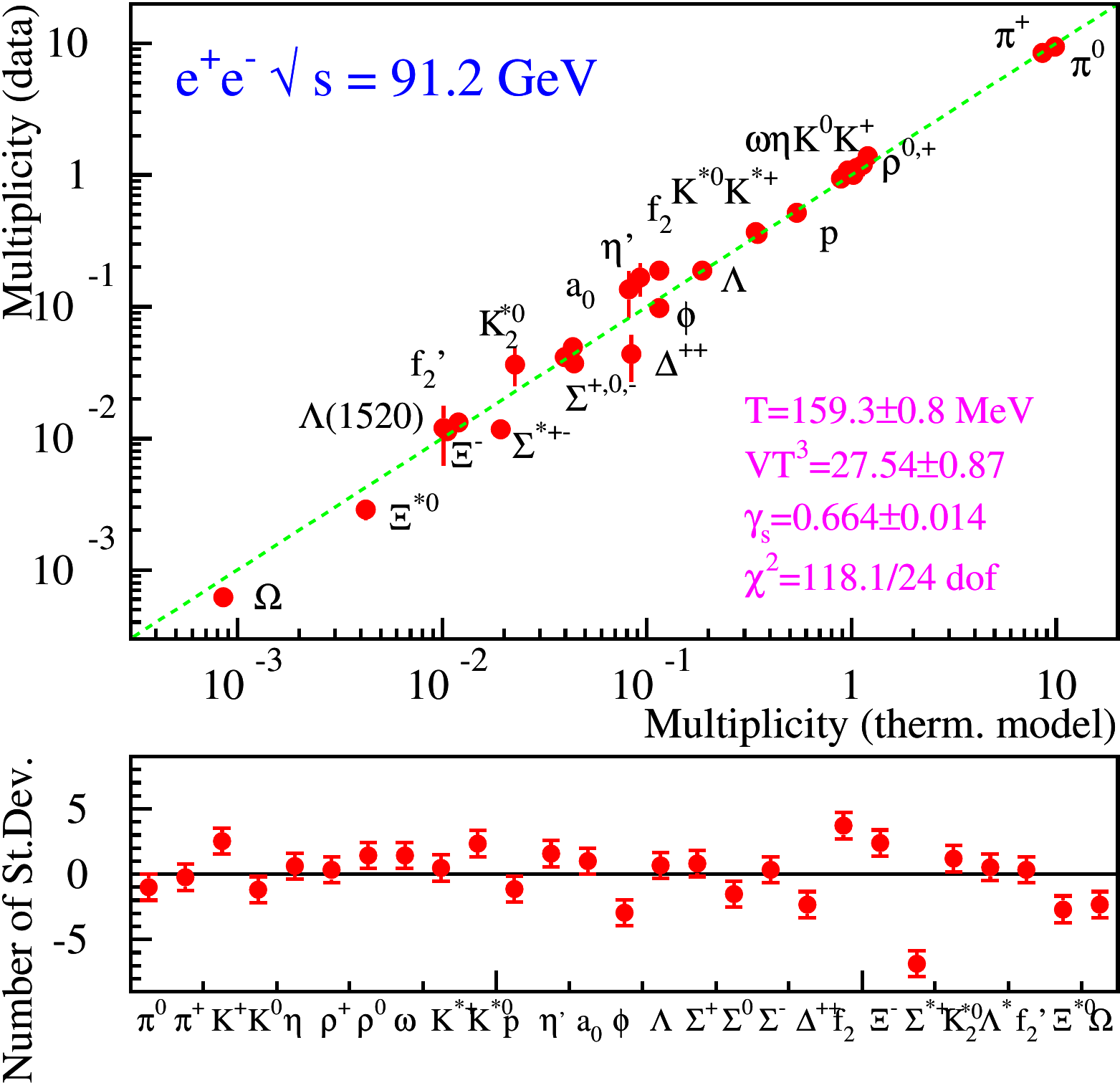}
%\picplace{5cm}{2cm} % Give the correct figure height and width in cm
%
\caption{Hadron production in electron-positron annihilation at LEP energy, 
confronted with canonical Statistical Hadronization Model calculations\cite{15}.}
\label{fig:1}       % Give a unique label
\end{figure}
We show in Fig.1 the LEP data at $91.2$~GeV confronted with the canonical SHM\cite{15},
revealing a hadronization temperature of 160 MeV. It agrees with Hagedorns
limiting hadronic temperature as was predicted, already in 1975, right with
the discovery of QCD, by Cabibbo and Parisi\cite{16}. The second line of argument 
concerning hadronization equilibrium stems from the recent idea\cite{17} to 
investigate the overlap between the Hadron Gas Model and lattice QCD with 
regard e.g. of the energy and entropy densities as a function of temperature.
\begin{figure}[t]
%\sidecaption
\center
\includegraphics[width=.6\textwidth]{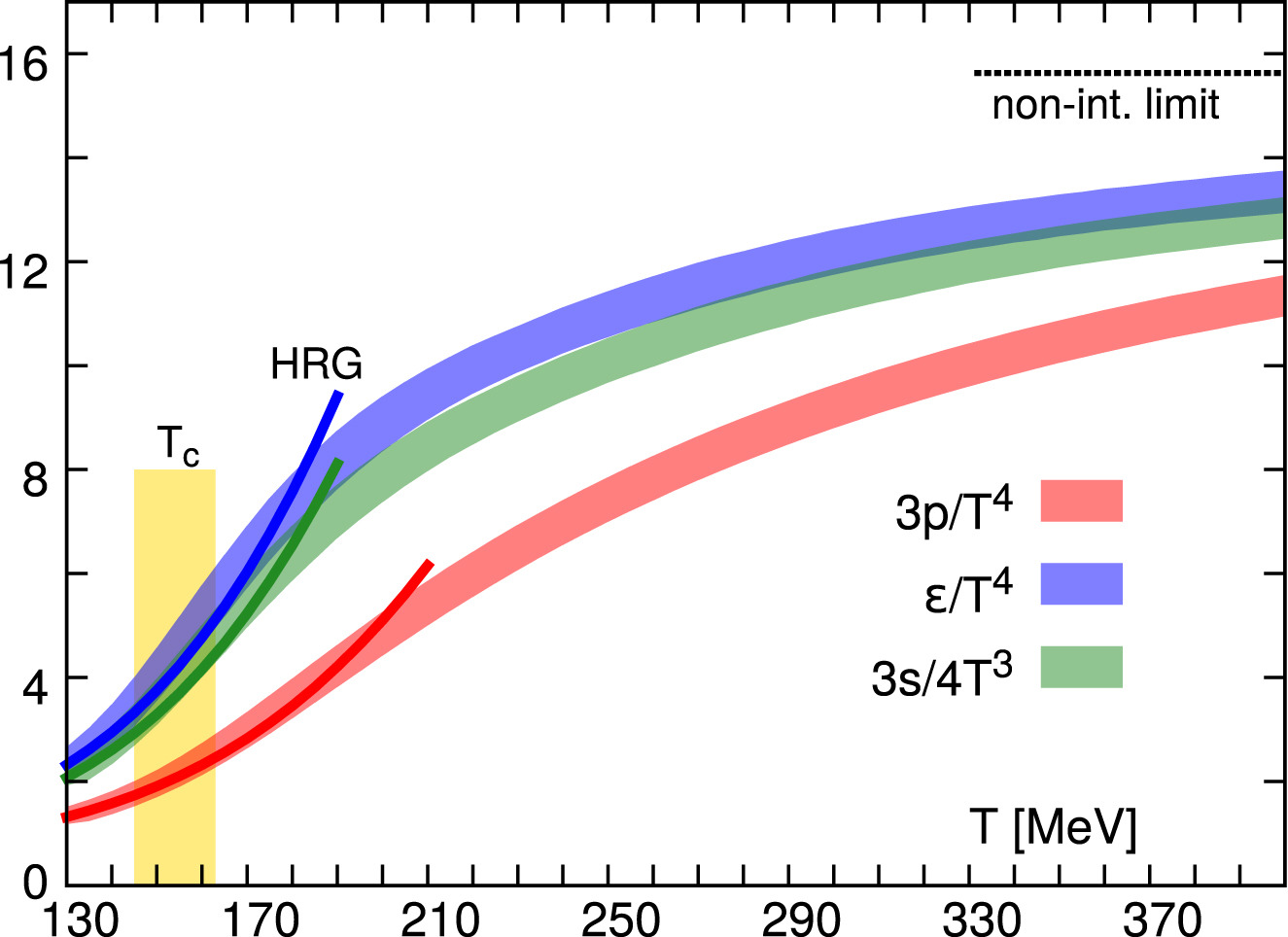}
%\picplace{5cm}{2cm} % Give the correct figure height and width in cm
%
\caption{Lattice QCD calculations for energy density and other quantities, 
confronted with Hadron-Resonance Gas model predictions for the same 
quantities\cite{18}.}
\label{fig:2}       % Give a unique label
\end{figure}

An example\cite{18} is shown in Fig.2 where we see the HRG curves intersect with the 
respective Lattice predictions in the vicinity of $165$~MeV. I.e. below this 
temperature the degrees of freedom of an equilibrium HRG appear to take
over from the lattice proper degrees of freedom. Note that the HRG and the SHM 
refer to the same hadron/resonance partition functions.

Next, we note that the results and predictions of lattice QCD are \textit{exact}(in 
the sense of the model) only at zero baryochemical potential where the phase
transition is a rapid cross-over. Toward finite $\mu_B$ various extrapolations
are employed\cite{2,3}, for example a Taylor expansion\cite{19} the coefficients of which 
are, in fact, related to experimentally accessible higher order fluctuations
of conserved quantities like baryon number and charge\cite{20,21}. We shall return 
to this work lateron but note, for now, that an interesting feature of the 
emerging phase boundary line at finite $\mu_B$ would be the intensely discussed 
critical point of QCD\cite{22,23}, at which the crossover nature of the 
parton-hadron transition would end giving way to a first order transition 
toward higher $\mu_B$. This would mark, on the one hand, the end of the lattice 
Taylor expansion, by divergence, but on the other hand lead to measurable 
changes in the sequence of hadronization points\cite{23} which are accessible via 
SHM analysis of the hadronic multiplicity distributions. This analysis could, 
therefore, not only ascertain the much debated existence of a critical point, 
but also serve to locate the phase transition line toward higher $\mu_B$ where 
lattice QCD is inapplicable, as of yet. 

The arguments above tacitly assume that the nucleus-nucleus collision dynamics 
does indeed cross the phase transformation line, and settles above it before 
re-expansion of the hot and dense matter volume sets in,
to approach the phase boundary. I.e. that we are at collisional energies above 
the so-called ``onset of deconfinement''\cite{24}. Estimates of the corresponding 
incident energy domain (in central, heavy nuclear collisions) remain uncertain 
at present, pointing to the $\sqrt{s}$ region from about $4$ to $8$~GeV. Below this 
(yet to be determined) energy the hadronic multiplicities would not stem from 
the QCD hadronization phase transition, which is responsible for the 
hadro-chemical equilibrium among the species' abundances\cite{4,5,9,10,11}.
Thus, unless another phase transition comes into play at high baryochemical 
potential\cite{25}, we should expect observable changes in the hadronic freeze-out 
pattern, such as a sequential chemical freeze-out in inverse order of inelastic 
cross section (characteristic of a diluting hadron gas). We shall return to 
this question at the end of this article but state, for now, that we can not 
yet arrive at clearcut conclusions concerning such features, due to a lack of 
data in the AGS energy domain, and below.

Turning to another important assumption of the above overall model 
considerations it was shown\cite{26,27} that the simple picture of an instant, 
synchronous chemical freeze-out of all hadronic species, occurring directly at 
the hadronization phase transformation, requires some revision.
Final state inelastic or annihilation processes need consideration owing to the 
high spatial particle density after hadronization. This effect is perfectly 
absent in elementary collisions such as  e$^{+}$--~e$^{-}$~annihilation to hadrons where 
hadrons are born into the physical vacuum. The SHM was initially developed for 
such elementary collisions\cite{28}. Turning to $A+A$ collisions it was shown that a 
detailed investigation of final state modifications of the multiplicity 
distribution exhibits relatively weak effects on the bulk meson production, 
pions and kaons which embody more than
$90\%$ of the total hadronic output energy at the LHC energy, but is important
via annihilation of baryons and antibaryons, notably p and pbar. These effects 
were quantified using the microscopic transport model UrQMD, in its
so-called \textit{hybrid} version\cite{29} where a hydrodynamic expansion mode of the
collisional volume is terminated by a simulation of the hadronization process
via the Cooper-Frye mechanism. The emerging hadrons and resonances are then
traced through the ensuing, final hadron-resonance cascade expansion phase of 
UrQMD, and its attenuating effects can thus be determined, species-wise.
These final state modifications affect the outcome of the SHM analysis as we 
shall demonstrate in the next section; they need to be taken into account 
before concluding on the position of the QCD phase boundary line.
Beyond studying the changes occurring in baryon and antibaryon multiplicity,
and their effects on the Statistical Hadronization model analysis\cite{26,27,30} as 
further discussed below, other groups have also focused on the consequences of
this final state annihilation for the hydrodynamic, notably the elliptic flow
of various identified hadronic species\cite{31}.

In the next section we shall discuss an approach to overcome the final state 
effects, by generating UrQMD-modification factors for the hadronic 
multiplicities that lead to a modified SHM analysis. Turning to SHM analysis of data at LHC,
SPS and AGS energies in Sec.~\ref{sec:3} we discuss the resulting new freeze-out curve\cite{30} 
and its differences to the result of a \textit{standard} SHM analysis, as well as to 
the often quoted smooth interpolation curve given by Cleymans \textit{et al.}\cite{32}. In 
Sec.~\ref{sec:4} we compare to the lattice QCD results for the parton-hadron phase 
boundary, confirming the lattice prediction\cite{2,3} that the line is almost 
horizontal up to a rather high $\mu_B$. We then turn to a discussion of the 
apparent tension between the SHM conclusion that $T_c=164 \pm 5$~MeV and the 
recent conclusion from consideration of data, HRG and Lattice QCD results 
concerning higher order fluctuations of proton and charge multiplicity 
distributions, that appear to favour a lower transition temperature, of about 
$150-155$~MeV\cite{33}. Section~\ref{sec:5} will give a short presentation of the issues concerning the SHM, i.e  acceptance 
and stopping power effects, the core-corona model, canonical strangeness 
suppression, and new LHC $p + p$ collision results. A discussion section will end 
the paper in Sec.~\ref{sec:7}, where we will briefly return to the issues of critical 
point, and onset of deconfinement.

It remains to say that we will not re-iterate the theoretical formulation of 
the grand canonical Statistical Hadronization model in this brief overview. 
Comprehensive presentations can be found in references \cite{4,5,6,8}.

\section{UrQMD final state effects and the Statistical Model analysis}
\label{sec:2}
Combined microscopic-macroscopic models are among the most promising approaches 
to describe nucleus-nucleus collisions \cite{34}. Initial state effects causing 
eventwise fluctuations of the dynamics can be installed and investigated 
separately, as is the case for the study of the ensuing hydrodynamic flow 
expansion, and for the matching to the final state hadron-resonance cascade 
that leads to decoupling. The latter aspect is our concern here.

In the hybrid UrQMD model employed\cite{26,27} to assess the final state effects on 
hadronic multiplicities in $A+A$ collisions, the hydrodynamic stage is a full 
(3+1) dimensional ideal hydro model executed by the SHASTA algorithm\cite{35}, 
employing the CH EOS from Ref.\cite{36} which corresponds to a crossover transition 
from a hadronic gas to the QGP at $\mu_B=0$ that extends to the region of finite 
$\mu_B$, relevant for all beam energies discussed, at present. The hydrodynamic 
evolution was stopped, generally speaking, once the energy density 
(temperature) in the system falls below a pre-set critical
value. Then the Cooper-Fry equations were sampled on a defined hypersurface, in 
accordance with conservation of all charges as well as the total energy. Two 
different choices of the hypersurface have been employed in order to assess the 
sensitivity of the ensuing \textit{afterburner} effects to the initialization 
procedure. The first choice was an effective iso-proper time hadronization\cite{37}, 
implemented by freezing out in successive transverse
slices, of thickness $dz = 0.2$~fm, whenever the last flow cell of the considered
slice fulfills the freeze-out criterion: an energy density $\epsilon$ below five 
times the nuclear ground state energy density (i.e. about $750$~MeV/fm$^3$).
This procedure was carried out separately for each slice, and the particle
vector information transferred to the cascade part of the UrQMD model\cite{29}. The 
effect of final state interaction was then
quantified by either stopping the calculation directly after hadronization,
letting the produced hadronic and resonance species undergo all their strong
decays as if in vacuum, thus establishing a fictitious multiplicity
distribution ideally referring to the hadronization \textit{point}. Alternatively,
the final afterburner UrQMD stage was attached, and the multiplicity 
distribution at decoupling  generated. For each hadronic species one could
then extract a modification factor indicating the strength of the afterburner
effects (a method introduced by Bass and Dumitru\cite{34}). This factor got finally 
employed in the SHM analysis of the data (see Sec.~\ref{sec:3}).

In a further investigation\cite{30} an alternative choice of the hadronization
hypersurface was explored, corresponding to an isothermal termination of the 
hydro expansion stage of UrQMD. Here, one implements the switch to 
hadronic/resonance degrees of freedom locally, in each hydro-flow cell,
once it falls below the pre-set energy density, or temperature,
calculating the hypersurface element with a state of the art
hypersurface finder\cite{38}.

In the following we illustrate this overall procedure, and its main 
consequences as far as the UrQMD predictions for final state attenuation of 
hadronic abundances are concerned. The resulting modification factors are 
employed in the data analysis with the Statistical Hadronization Model(Sec.~\ref{sec:3}). 
For illustration of the phenomena we choose the case of central $Pb+Pb$ 
collisions at the top SPS energy, $\sqrt{s} = 17.3$~GeV (one of the data sets to 
be analyzed in Sec.~\ref{sec:3}).
\begin{figure}[t]
%\sidecaption
\center
\includegraphics[width=0.8\textwidth]{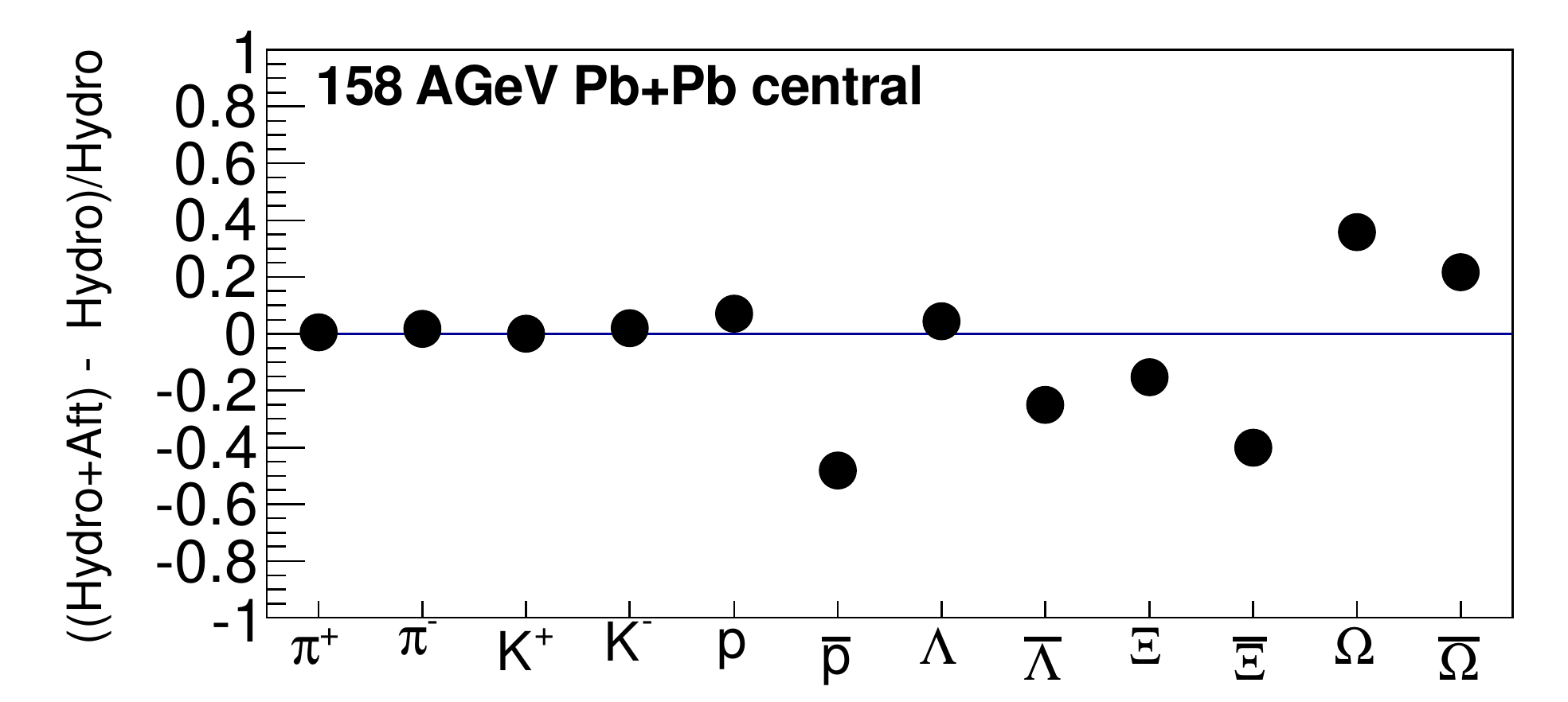}
%\picplace{5cm}{2cm} % Give the correct figure height and width in cm
%
\caption{Modification factors accounting for final state attenuation in the 
hadron/resonance expansion\cite{26,27}.}
\label{fig:3}       % Give a unique label
\end{figure}
Figure~\ref{fig:3} shows the modification factors\cite{26}
\begin{equation}
    M = \frac{yield\text{(with\;afterburner)} - yield\text{(at\;hadronization)}}{yield\text{(at\;hadronization)}}\;
\label{equ:1}
\end{equation}
for the hadronic species covered by the experiment. One sees the pion, kaon,
proton, Lambda and Xi yields, i.e. the bulk output from hadronization, to be 
essentially unaffected at this incident energy, a remarkable result.
Thus far the assumption of a synchronous freeze-out, directly at hadronization,
as made in the standard version of the SHM\cite{4,5,6,7}, is substantiated. On the
other hand, antiprotons suffer a net loss of about 50\%, and Anti-Lambda and 
Anti-Xi multiplicities are reduced by $25$ to $40\%$. The Omega/Anti-Omega hyperons are
less affected because they have no excited resonant states, and freeze out
from the cascade evolution almost instantaneously, owing to their low total
cross section. There may also occur contributions of baryon/antibaryon
regeneration\cite{39} which are only partially implemented in UrQMD. A new study of 
their importance is under way with the QHDS transport model\cite{40}.

We see that in the UrQMD cascade evolution inelastic reaction channels are of 
no major importance whereas baryon-antibaryon annihilation channels do not 
freeze-out directly after hadronization. At the relatively low SPS energy, 
illustrated here, baryons are far more abundant than their anti-partners; thus 
their partial annihilation with antibaryons changes a minor fraction of their 
multiplicities, whereas the antibaryons are significantly diminished, 
fractionally. This pattern changes at LHC energies due to the near-perfect
particle-antiparticle symmetry, prevailing there\cite{27}. Under such conditions the 
antibaryons are equally affected.

These attenuations have a profound effect on the Statistical Model analysis. 
The loss of antibaryons is, clearly, a departure from the initial chemical 
equilibrium distribution which is reflected in a SHM analysis (described in 
Ref.\cite{26}) of the two sets of UrQMD multiplicities, obtained with, and without the 
afterburner stage.

\begin{figure}[t]
\centering
%\sidecaption
\begin{minipage}[b]{0.49\linewidth}
	\includegraphics[width=\textwidth]{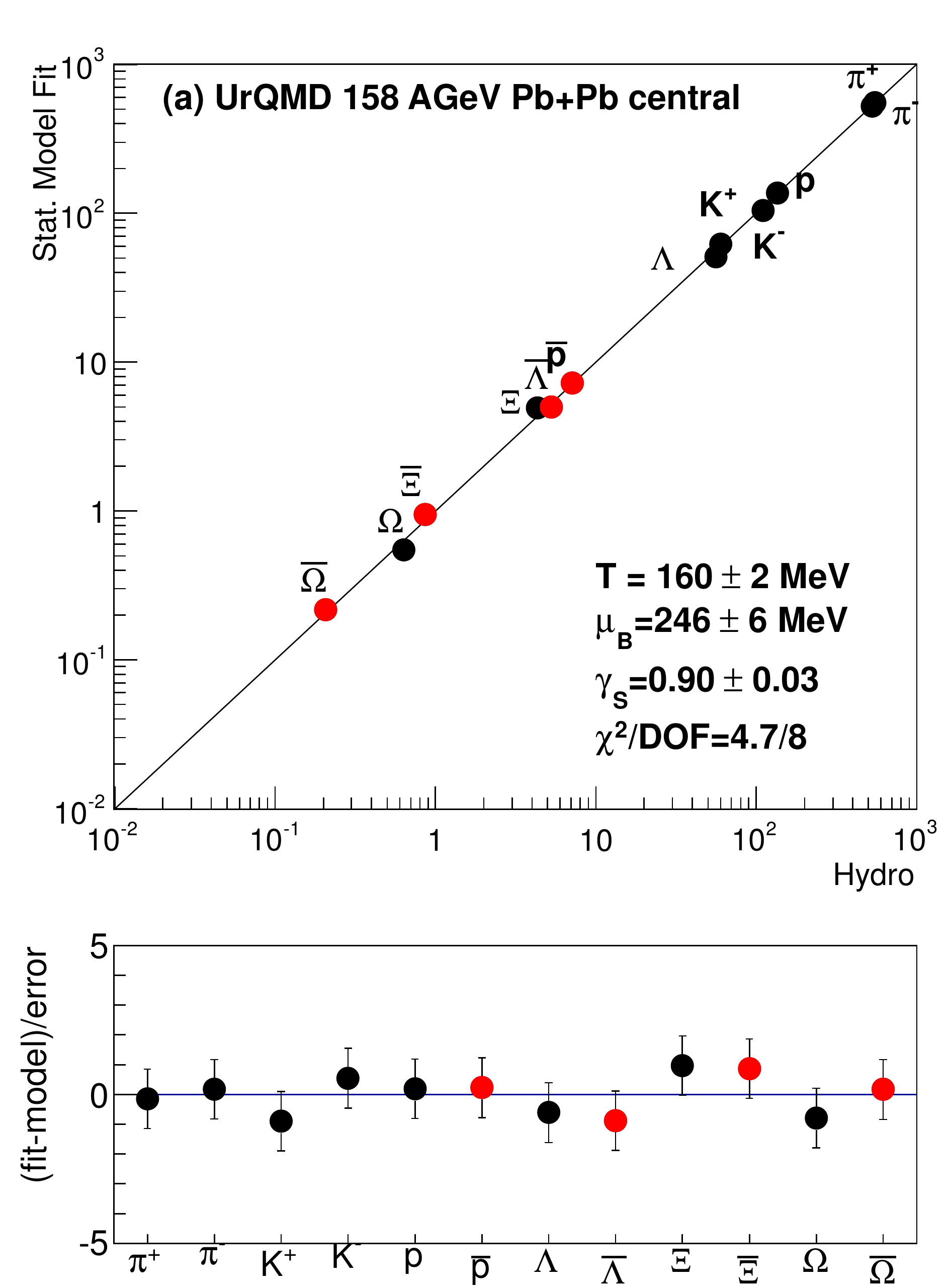}
\end{minipage}
\begin{minipage}[b]{0.49\linewidth}
	\includegraphics[width=\textwidth]{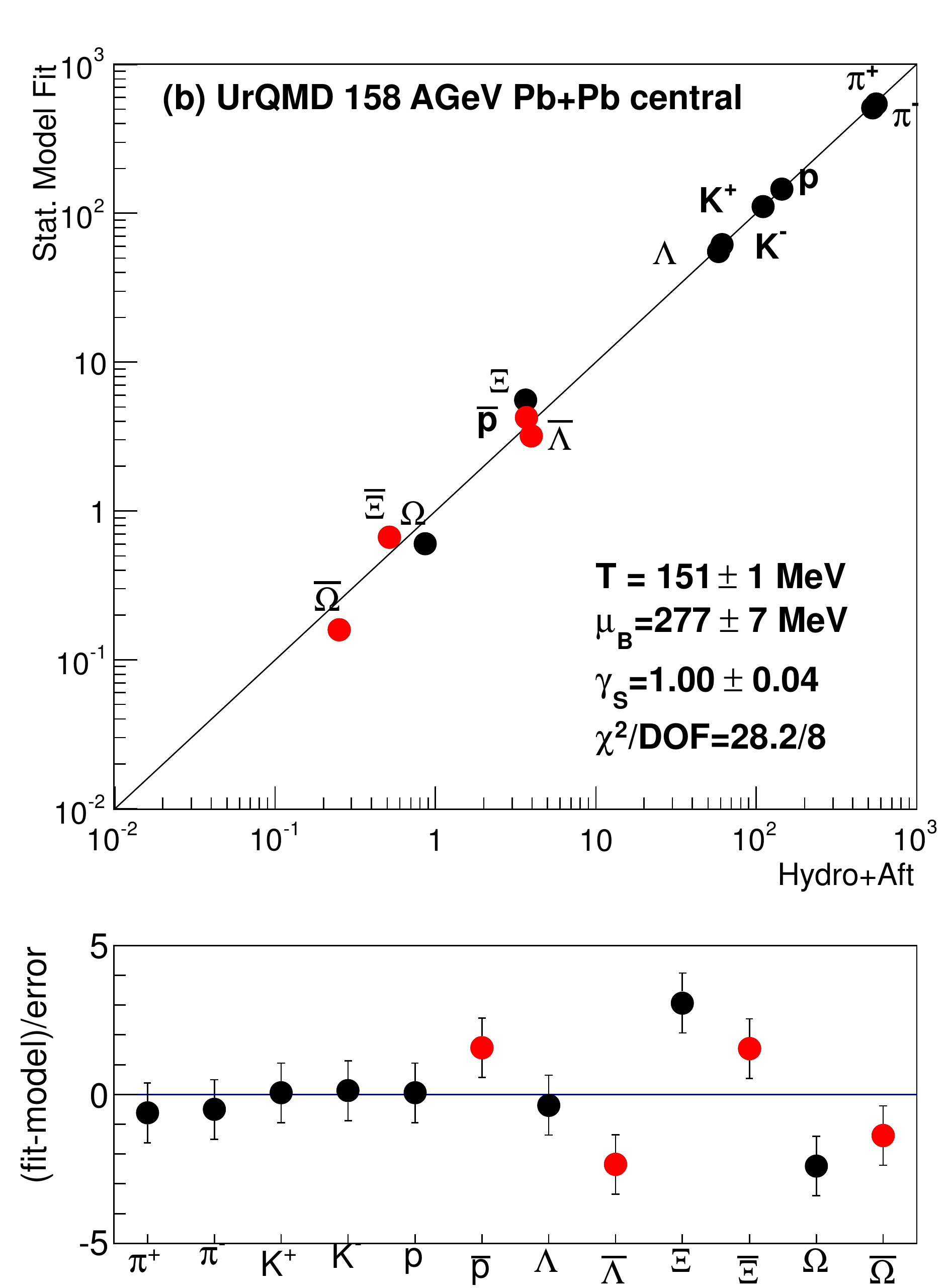}
\end{minipage}
%\picplace{5cm}{2cm} % Give the correct figure height and width in cm
%
\caption{UrQMD multiplicities fitted by the 
Grand Canonical model, fitted to the ``data'' right after hadronization, and to 
the results with attached UrQMD afterburner\cite{26}.}
\label{fig:4}       % Give a unique label
\end{figure}
This is shown in Fig.\ref{fig:4}: the UrQMD multiplicities, taken as 
``data'', have been fitted with the SHM. We see that, on the one hand, the SHM 
fit to the no-afterburner results is of excellent quality and reveals a 
hadronization temperature of 160 MeV. This is not surprising because the SHM
fit merely responds to the chemical equilibrium multiplicity distribution
imprinted by the Cooper-Frye hadronization, with its preset temperature value.
On the other hand, the case with afterburner shows a very much deteriorated
fit in the antibaryon sector, at a temperature reduced to $151$~MeV. The apparent 
chemical freeze-out temperature thus drops down, significantly, due to the 
final state distortions of the multiplicity distribution. The
traditional statistical model analysis thus needs revision. We will show this
in the next section, turning to data analysis.

\section{Data Analysis: the SHM Freeze-out Curve}
\label{sec:3}
In this section we illustrate the UrQMD-modified Statistical Model analysis\cite{30} 
that addresses 5 hadronic multiplicity data sets obtained, respectively, by 
ALICE\cite{41} at the LHC energy of $2.76$~TeV per nucleon pair, by NA49\cite{42} at the 
SPS energies $17.3$, $8.7$ and $7.6$~GeV, and by the AGS experiments E891, 802, 896,
877\cite{43} at $4.85$~GeV. All data refer to central collisions of $Au+Au$ (at the AGS)
and $Pb+Pb$ (at SPS and LHC). Note that the new STAR data\cite{44} from the RHIC BES 
program have not been included in this analysis because they are not yet 
corrected for antiproton feed-down from secondary weak antihyperon decays.
\begin{figure}[t]
%\sidecaption
\includegraphics[trim=1.2cm 1.2cm 1cm 1.5cm, width=.48\textwidth]{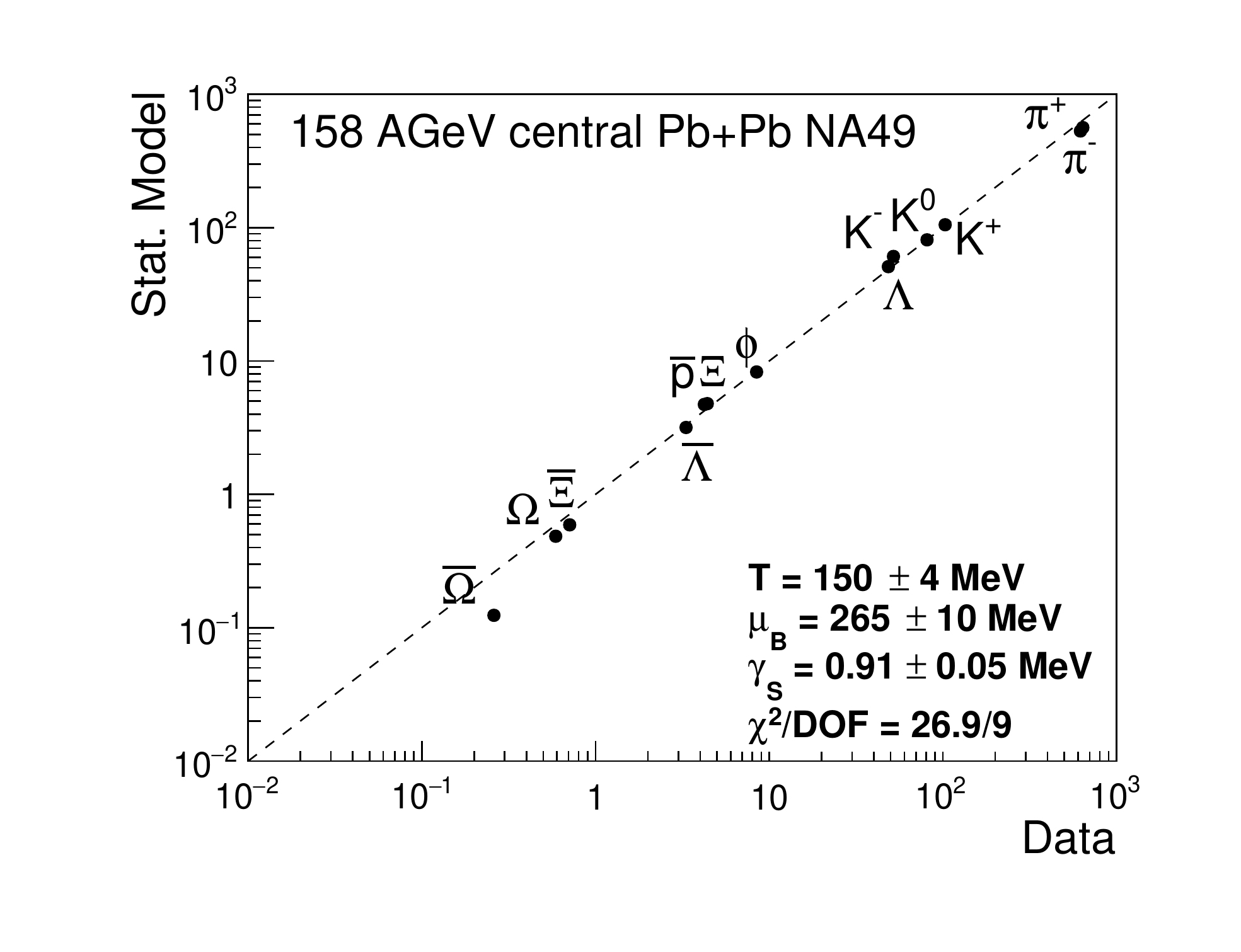}
\quad
\includegraphics[trim=1.2cm 1.2cm 1cm 1.5cm, width=.48\textwidth]{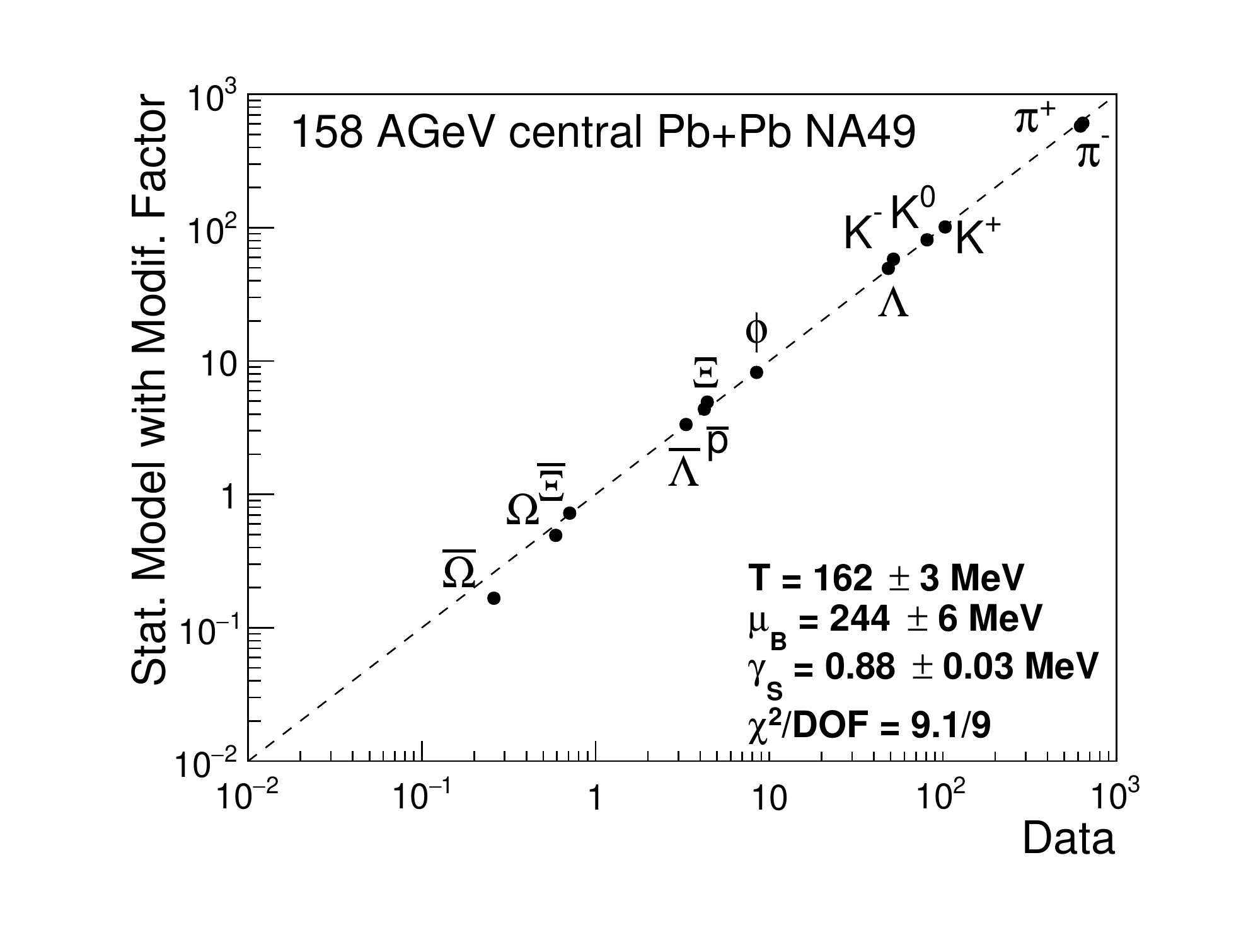}
%\picplace{5cm}{2cm} % Give the correct figure height and width in cm
%
\caption{Hadron multiplicities from central $Pb+Pb$ collisions at top SPS energy, 
confronted with HRG fits employing the standard Grand Canonical approach, and 
the SHM corrected by UrQMD modification factors\cite{26}.}
\label{fig:5}       % Give a unique label
\end{figure}
In Fig.~\ref{fig:5} we illustrate this analysis choosing the case of the top SPS energy 
data, central $Pb+Pb$ collisions at $17.3$~GeV. The outcome of the \textit{traditional} 
SHM analysis\cite{6} is compared with the version in which the SHM fit to the data 
is done with theoretical multiplicities modified by the final
state attenuation factors from UrQMD (see Fig.\ref{fig:3}). Two major effects are 
obvious: the deduced temperature increases significantly, from $T = 150$~MeV in 
the former, to $T = 162$~MeV in the latter case. Concurrently, the fit 
chi-square/dof drops dramatically, from about $2.7$ to about $1.0$. The 
UrQMD-reconstructed yield distribution, which aims to re-establish the
distribution at hadronization, obviously meets with the grand canonical
equilibrium hypothesis, implicit in the SHM approach. Recalling the
analogous observations derived above, from the SHM study of UrQMD
predictions for the multiplicity distribution (Fig.4), we re-iterate
the conclusion that the final state hadron/resonance cascade evolution (here 
represented by the UrQMD \textit{afterburner}), generates a
distortion of the hadro-chemical equilibrium yield distribution that is 
initially imprinted at hadronization. This is reflected in the
unsatisfactory SHM fit quality, a feature also reported from the SHM
analysis of Andronic \textit{et al.}\cite{12}.

A short remark is in order here, to clarify the above use of the term 
equilibrium. First, it refers, not, to a global equilibrium of all aspects of 
the phase space distributions of the hadrons that emerge from the QCD 
hadronization phase transformation. It refers, only, to its first moment,
the hadronic yield (multiplicity) distribution. Higher moments, such as 
collective, discrete hadron emission flow patterns (radial, directed, elliptic 
and higher flow orders) represent non-isotropic momentum space flow moments at 
any higher order, in the final state. Furthermore, if the hadronization process 
creates a hadronic species thermal equilibrium distribution (which is 
represented by the quasi-classical equilibrium grand canonical Gibbs ensemble 
underlying the Statistical Model) directly corresponding to the conditions 
prevailing at hadronization (chiefly the energy density/temperature), this 
equilibrium state, if it freezes-in directly at this point (as assumed in the 
traditional SHM analysis), is of course out of equilibrium immediately after the 
onset of expansive dilution and cooling. The yield distribution does, ideally, 
not adjust to the falling energy density but stays, frozen into the ongoing 
expansion, as we have seen in Fig.\ref{fig:3}, to appear as the bulk hadronization
output. The hadron/resonance expansion stage will adjust other, higher
moments. We note that also the higher moments of conserved baryon number and
charge multiplicity fluctuation may not become stationary (freeze out) at the
very instant at which the number and entropy densities of the hadron/resonance 
degrees of freedom get fixed. We shall return to this consideration below, when 
discussing the recent work on fitting kurtosis and skewness data, and HRG 
predictions, to the corresponding results from lattice QCD. Thus there are two, 
or perhaps even three successive freeze-outs: the primordial hadronic species 
distribution reflects the energy/entropy densities prevailing at the onset of 
hadronization (with corrections that this study quantifies) whereas higher 
order multiplicity fluctuations may be imprinted throughout the
duration of the hadronization period (which is finite in a cross-over 
transition). Finally, spectral and correlation features become stationary at 
the lower temperatures of decoupling from all strong interaction(kinetic 
freeze-out).

\begin{figure}[b]
%\sidecaption
\centering
\includegraphics[scale=.42]{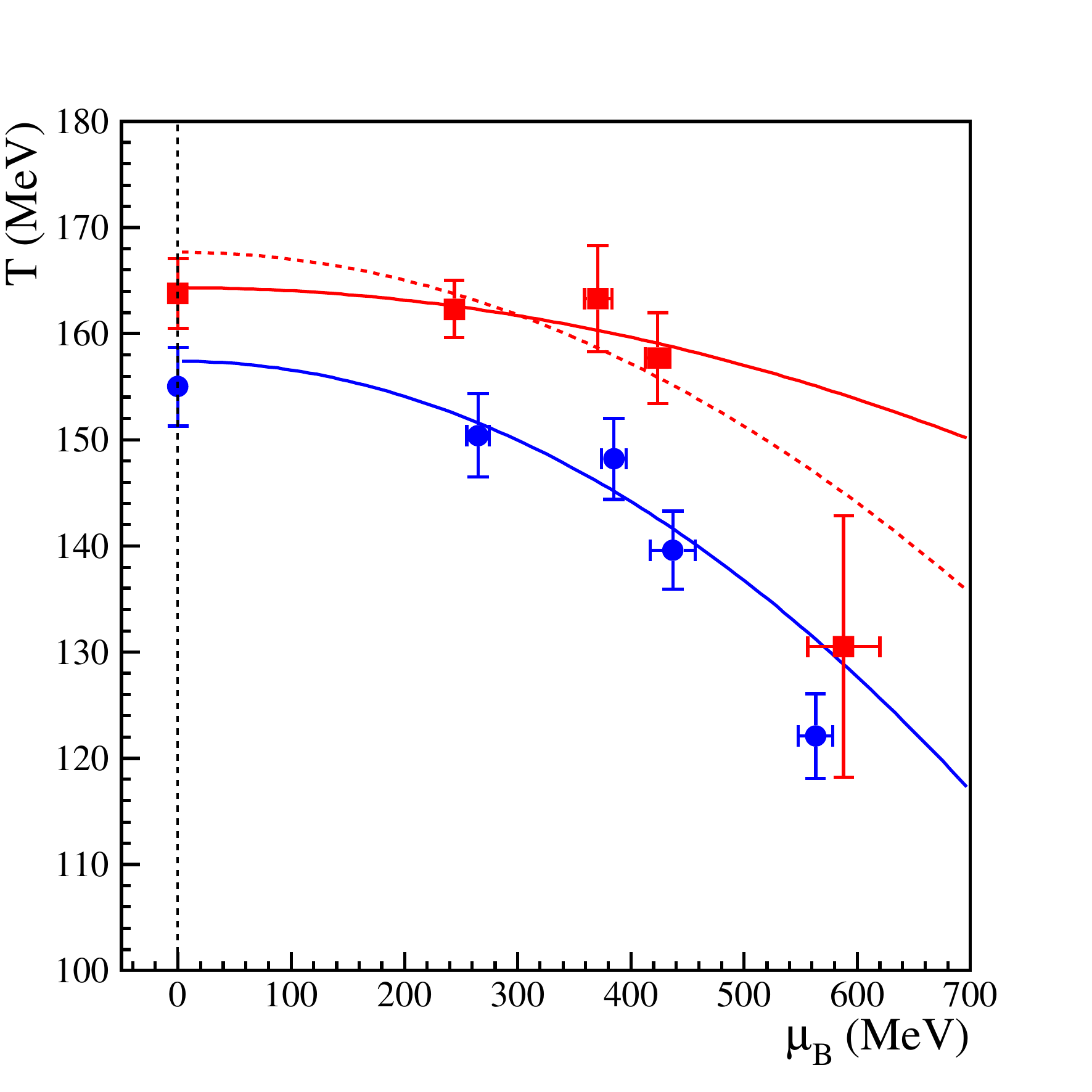}
%\picplace{5cm}{2cm} % Give the correct figure height and width in cm
%
\caption{Results of the UrQMD plus SHM approach, for hadronization points in the 
($T$,$\mu_B$) plane. The lower set of points represents a standard grand canonical 
SHM analysis\cite{30}.}
\label{fig:6}       % Give a unique label
\end{figure}

The results from Ref.\cite{30} concerning the reconstruction of the hadronization 
line, in the plane of temperature $T$ and baryochemical potential $\mu_B$ are 
summarized in Fig.~\ref{fig:6}. 
It shows the ($T$,$\mu_B$) points at the 5 energies considered 
here, that result from SHM analysis of the corresponding hadron
multiplicity data sets. The results of the standard SHM approach are
compared to the UrQMD-modified SHM in both the hadronization modes described
in Sec.~\ref{sec:2}, isothermal and slice-by-slice. The latter give almost
indistinguishable results pointing to the conclusion that the detailed
method employed for UrQMD hadronization has little influence on the
effects of the ensuing cascade expansion stage, as far as multiplicities
are concerned. Up to a baryochemical potential of about $370$~MeV the points 
exhibit very little downward slope, unlike the results from standard SHM 
analysis which drop off more steeply. This invites comparison to the lattice 
QCD predictions for the curvature of the phase boundary line, a subject of the 
next section, as is the apparent abrupt downward turn toward higher 
$\mu_B$ (unfortunately documented by a solitary AGS point only). As
far as the turn-off toward higher baryochemical potential is concerned,
which may indicate the advent of new physics as discussed in the next
section, we note here that this feature has been well recognized already
in previous investigations of the SHM freeze-out, which did not employ any
final state attenuation corrections.
\begin{figure}[t]
%\sidecaption
\centering
\includegraphics[scale=.7]{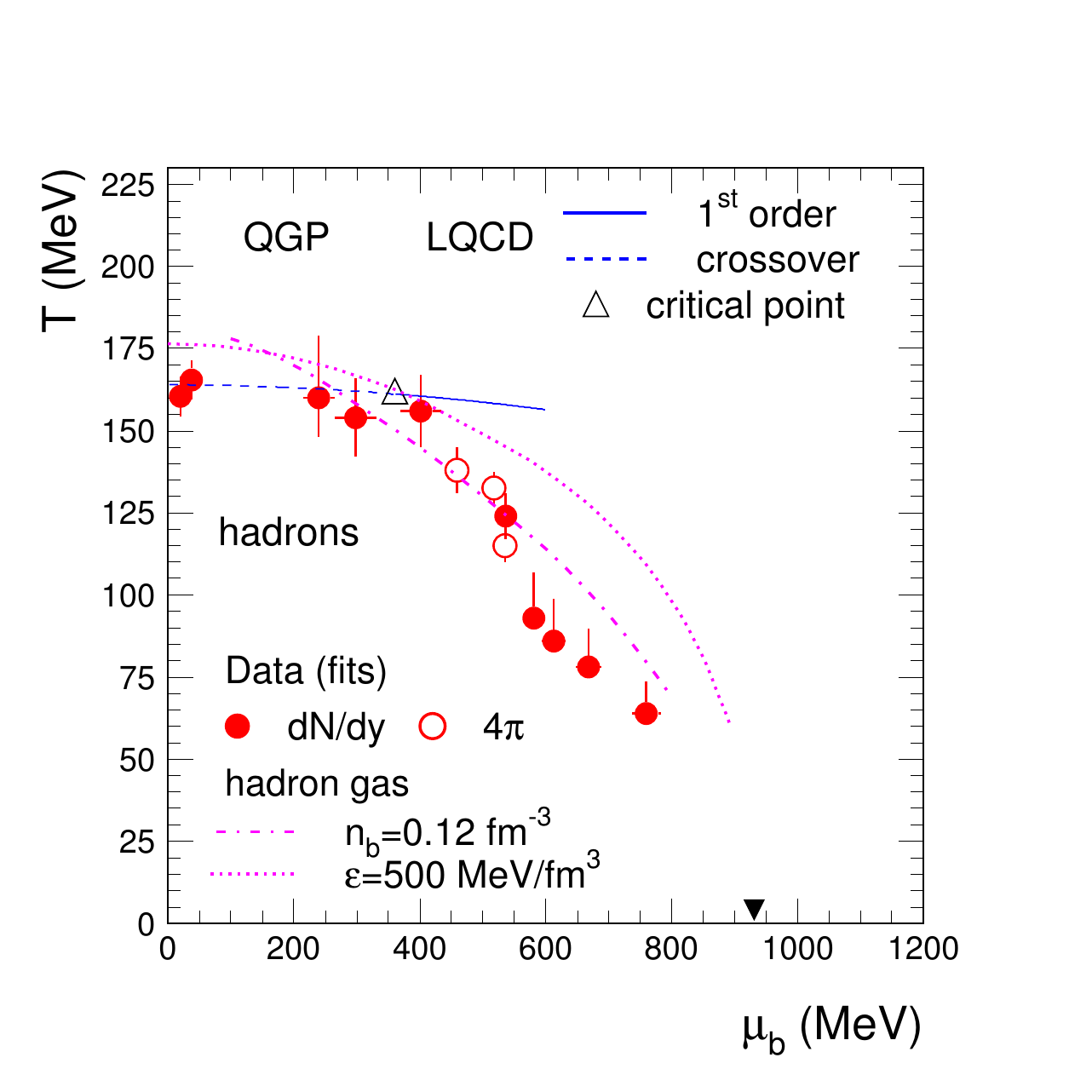}
%\picplace{5cm}{2cm} % Give the correct figure height and width in cm
%
\caption{Standard SHM analysis results for freeze-out points in the (T,$\mu_B$) 
plane, covering energies from RHIC, via the SPS and AGS, down to SIS18\cite{7}.}
\label{fig:7}       % Give a unique label
\end{figure}
We show in Fig.\ref{fig:7} the state of the art as of 2006 by reproducing the results obtained by Andronic \textit{et al.}\cite{7} for the SHM freeze-out systematics as function of $\mu_B$, gathering all results
available at that time. Indeed, the authors note that the turn-off,
corresponding to SHM investigations below a CM energy of about $8$~GeV,
is hardly compatible with any smooth interpolation.

Finally, we note here that the much discussed puzzle\cite{12,41} of an apparent 
\textit{non-thermal}, high pion to proton ratio observed in the ALICE data\cite{41} at the 
LHC energy of $2.76$~TeV (which resulted in an unsatisfactory SHM fit\cite{12}) finds 
its resolution in the UrQMD modified approach illustrated above. Final state 
annihilation of nucleons and antinucleons reduces the p and pbar multiplicities 
while increasing the pion yield as each annihilation contributes 5 pions, on 
average. Both the p, pbar losses, and the additional pion gains are, of course, 
of non-thermal origin.

\section{Comparison to Lattice QCD}
\label{sec:4}

We wish to discuss, in more detail, the observations described above.
\begin{alphlist}[]
\item The small initial slope of the UrQMD modified SHM model points resembles the
recent lattice predictions for the QCD transition line\cite{2,3} which employ a
quadratic ansatz,
\begin{equation}
T_{c,\mu_B} =  T_{c,0} ( 1 - \kappa_{2,B})(\mu_B/T_{c,0})^{2}
\label{equ:2}
\end{equation}
reporting curvature values $\kappa_{2,B}$ between $0.007$  in Ref.\cite{2} and $0.015$ in Ref.\cite{3}. The first 4 
points are compatible with $T_{c,0} = 164$~MeV and $\kappa_{2,B} = 0.0048$.

\item The steep drop-off in the $\mu_B$ domain above about $400$~MeV may represent a 
manifestation of a fourth order coefficient  $-\kappa_{4,B}(\mu_B/T_{c,0})^{4}$  %-kappa4(B)($\mu_B$/T(c,0)**4
in the lattice Taylor expansion. This, then, would be of order $0.003$ but we note that 
this estimate is based on a single data point (AGS) only. We can not decide, 
with certainty, whether the QCD phase boundary moves downward, steeply, in this 
domain. Alternatively, the steep drop-off could signal a change in the 
collisional evolution, such as the advent, at high $\mu_B$, of the hypothetical 
quarkyonic matter phase\cite{25} interpolating between the domains of deconfined, 
and of hadron-resonance matter in the QCD phase diagram, thus
shifting the final hadronization transition downward in temperature. 
Hadronization from a quarkyonic matter phase has not yet been considered 
theoretically so that we can not say with certainty that it would feature a 
similarly phase space dominated yield distribution over species. Finally, and 
more simply, we might witness the so-called ``onset of deconfinement''\cite{24} at 
the AGS energy, where the collisional volume does deconfine only partially,
thus lacking a phase transition synchronized hadronic freeze-out.

\item Obviously, the reconstructed hadronization points in Fig.~\ref{fig:6} indicate a 
\mbox{(pseudo-)critical} temperature $T_{c,0}$ of $164\pm5$~MeV, at $\mu_B = 0$ and, likewise,
a $T_{c}$ above about $160$~MeV in the entire $\mu_B$ interval up to about $370$~MeV.
This observation is, clearly, at odds with the recent consensus in lattice QCD 
studies about a lower $T_{c,0}$ value, in the vicinity of $150$ to $155$~MeV\cite{33}. 
However, let us not rush for conclusions here. As we said above, higher order 
fluctuations (susceptibilities) might freeze out later than the zero-order 
multiplicity quantities (their overlap with lattice QCD thus occurring at a 
lower T) and, moreover, they might not be reliably accounted for by the 
noninteracting HRG model due to attractive short range interactions\cite{45}. Thus 
the HRG-Lattice overlap temperature, as derived from the
fluctuation/susceptibility observables, may not be a good approximation of 
$T_{c,0}$. Finally we recall that Fig.~\ref{fig:2} did suggest a $T_{c,0}$ above $160$~MeV
from state of the art lattice calculations for the equation of state and the 
energy density --- the quantities more closely related to hadron/resonance number 
density (that is of relevance in the SHM and also, of course, in the HRG) than 
e.g. the Polyakov-loop slope turnover, or higher susceptibilities. Addressing 
the former observables the HRG-lattice overlap region rather suggested a $T_{c,0}$ 
above $160$~MeV.

\item The UrQMD-modified results represent a revision of the \textit{hadronic freeze-out
curve} that was obtained\cite{32} by an interpolation of the ($T$,$\mu_B$) values 
deduced from grand canonical SHM analysis of central $A+A$ collisions, gathered 
from RHIC energies down to AGS and SIS18 energies\cite{4,5,6,7}. This often
shown curve also extrapolated to a $T_{c,0}$ of about $165$~MeV. However, the
various sets of hadron multiplicity data, available by 2006 for AGS, SPS and 
RHIC energies, had mostly not yet been corrected for experimentally unresolved 
feed-down contributions stemming from weak decays of the hyperons/antihyperons. 
This increased the apparent multiplicities for
baryons/antibaryons, reflected, in the SHM analysis, by too high temperature 
assessments. The data sets used in Fig.~\ref{fig:6} are properly corrected, thus the 
corresponding standard freeze-out curve moves
below the 2006 result, by about $10$~MeV as shown in Fig.~\ref{fig:3}. We note that,
anyhow, this standard SHM result looses interest due to the annihilation 
processes in the final cascade evolution, which have, first of all, to be 
accounted for, by an analysis like the one described above. Whether or not
the particular UrQMD approach represents the last word about this issue
clearly remains to be clarified.
\end{alphlist}

\section{Issues of Concern in SHM Analysis}
\label{sec:5}

Let us begin this chapter with a general observation. The consistency of 
existing SHM analyses suffers  from the plurality concerning details of the 
employed analysis formalism or, even, its basic formulation. We shall take up 
some such aspects in some more detail below but mention, for now, the options 
for or against e.g. hadron eigenvolume corrections, strangeness correlation 
volumes or a strangeness fugacity, corona-core separation, light quark 
fugacity, to name just a few. Furthermore, the available data reflect different 
experimental acceptance conditions (both in rapidity space and in azimuthal 
coverage), and data sets comprise a varying sub-selection of hadronic species, 
the latter often a consequence of hyperon multiplicities falling strongly with 
incident energy. Thus, a set of multiplicities ranging from pions up to 
anti-Omega hyperons gets employed from LHC down to top SPS energies, but the 
rarer species fall away, gradually, toward lower energies. This is 
counterproductive if one tries to secure a systematic coverage of the overall 
$T$, $\mu_B$ dependence. Overall, we note that the ideal view represented by the 
standard grand canonical ensemble may not be universally congruent to the
reality of $A+A$ collisions\cite{6}, and of their measurement within certain fixed
rapidity windows (see below). Thus the data sets of multiplicities may reflect
\textit{tensions} that vary with energy; the acceptance problem, for example,
disappears only with approaching boost invariance, i.e. at top LHC energy.
And the annihilation corrections affect, significantly, only the antibaryons at low energy because of the large excess of B over Bbar. Whereas both B and Bbar are attenuated at high energies.

\begin{figure}[t]
%\sidecaption
\centering
\includegraphics[scale=.3]{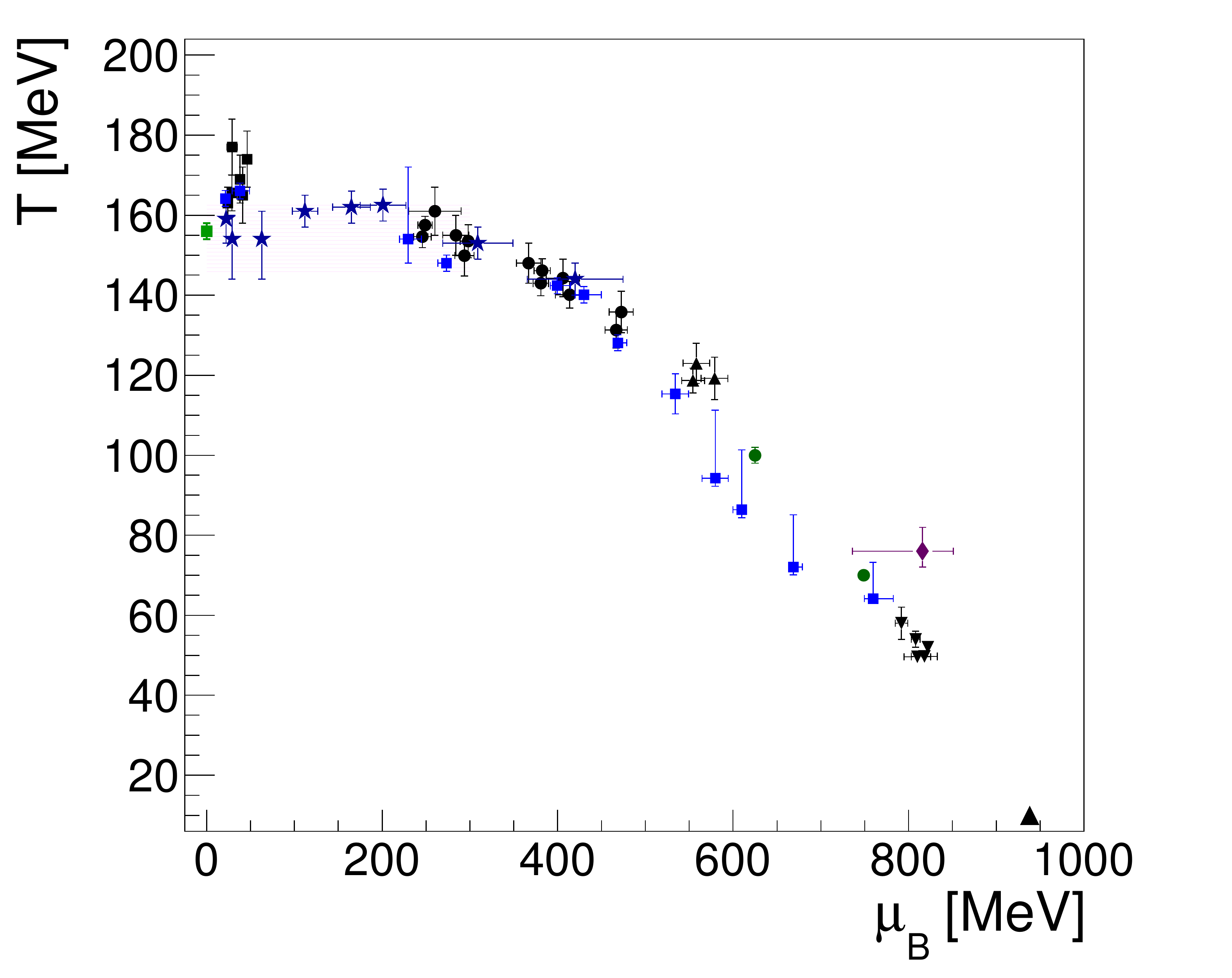}
%\picplace{5cm}{2cm} % Give the correct figure height and width in cm
%
\caption{Summary plot of present SHM analysis with various variants of the 
Statistical Hadronization Model, for the overall freeze-out curve in ($T$,$\mu_B$) 
space\cite{46,47}.}
\label{fig:8}       % Give a unique label
\end{figure}

Numerous further SHM analysis problems can be enumerated. An example is the 
question of how to treat the Phi-meson: with strangeness suppression (creation 
from K$^{+}$--~K$^{-}$ pairs) or without(strangeness zero at hadronization).  A 
further, new question: should the charmed hadrons be included in the general 
SHM fit(with a special fugacity factor as charm is not thermally produced), or 
should we expect a sequential hadronization of the heavier quarks\cite{46}? We shall 
take up some of such questions below but end here showing a fairly complete 
selection of available standard SHM analyses\cite{47}, from LHC down to SIS18 
energies, in Fig.~\ref{fig:8}\cite{48}.
 Quite expectedly the points exhibit considerable 
dispersion even though the results with UrQMD modification(Fig.~\ref{fig:6}), and the 
predictions of the Single Freezeout Model\cite{48} have not been included here.

\subsection{Acceptance and stopping power effects in rapidity space}
\label{sec:6.1}

The thermal rapidity spreading width of a single midrapidity fireball is 
proportional to $(T/m)^{1/2}$ for a hadron species of mass $m$\cite{49}. For pions 
emitted at $T = 160$~MeV this amounts to the interval $-1.3 < y < 1.3$, whereas 
emitted Omega hyperons would occupy a much smaller interval, of about $-0.4 < y 
< 0.4$. Kaons fall inbetween. Realistically, a fireball of $160$~MeV temperature 
could get created at top SPS energy (recall Fig.~\ref{fig:6}), but here it features, in 
addition, collective radial flow expansion which elongates the longitudinal 
rapidity distributions\cite{50}. The pions thus acquire a width of about $\pm1.5$ 
units of rapidity whereas all heavier hadrons stay within an interval of unit 
rapidity. This is the STAR experiment acceptance, employed in the BES RHIC runs 
at low energy, down to the SPS regime where considerations such as above may be 
realistic. Such an acceptance thus creates an artificial strangeness
enhancement, as reflected in the so-called Wroblewski ratio\cite{51}. It counts
the total number of s and sbar quarks, relative to the light quark number,
recovered in the observed hadronic multiplicities. The pions that carry the
major fraction of the light quarks partially leak out of the narrow
acceptance. This creates \textit{tensions} in the SHM analysis.

\begin{figure}[t]
%\sidecaption
\centering
\includegraphics[width=.95\textwidth]{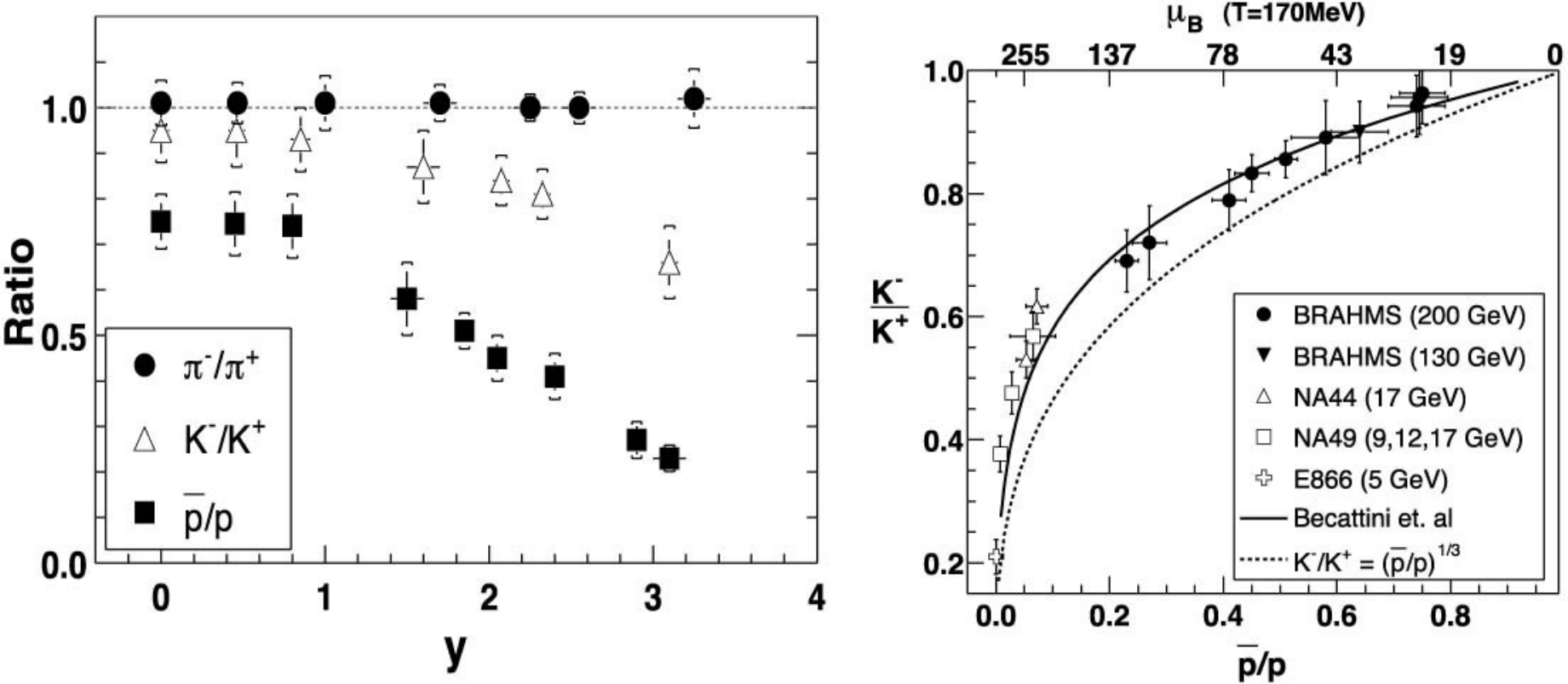}
%\picplace{5cm}{2cm} % Give the correct figure height and width in cm
%
\caption{Results of the RHIC BRAHMS experiment\cite{53} for antihadron to hadron ratios 
at top RHIC energies. The left panel shows the rapidity dependence, the right 
confronts a correlation plot of K$^{-}/$K$^{+}$ vs. pbar/p with SHM calculations at 
constant temperature but with decreasing baryochemical potential.}
\label{fig:9}       % Give a unique label
\end{figure}

A remedy might be to widen the experimental acceptance. The resulting effects 
can be studied with data from fixed target experiments, such as NA49 at the 
SPS, and BRAHMS at RHIC. The former already reported a minimum at midrapidity 
in the net baryon rapidity distribution for central $Pb+Pb$ collisions at $17.3$~GeV c.m. energy\cite{52}. This signals a minimum of the baryochemical potential 
which, in turn, is a consequence of the so-called stopping power of hadronic 
matter. The net baryon number density reflects the final distribution, in 
longitudinal phase space, of the valence quarks carried in by the colliding 
nuclei. In a Glauber-like picture the transport of valence quarks, from initial 
target/projectile rapidity positions inward toward midrapidity occurs via the 
successive binary collisions of nucleons (or their remnants) during the 
primordial interpenetration phase, which extract partons from the initial 
nucleon structure functions. This stopping mechanism spreads the bulk of the 
valence quarks over about $2.5$ units of rapidity in central mass $200$ nuclear 
collisions. From top SPS energy onward, the rapidity gap between the initial 
target and projectile rapidities, increases beyond $dy = 6$, and the midrapidity 
region is increasingly devoid of valence quarks, thus $\mu_B$ falls toward zero 
here. This can be perfectly illustrated by the BRAHMS results\cite{53} shown in 
Fig.~\ref{fig:9}.
The left panel shows results for central $Au+Au$ collisions at 200 GeV, 
exhibiting the rapidity dependence of the multiplicity ratios $\pi^{-}/\pi^{+}$, 
K$^{-}/$K$^{+}$ and pbar/p.
The latter two are proportional to $\exp(-2/3\; \mu_B/T)$ and $\exp(-2\; \mu_B/T)$, 
respectively, thus presenting a direct measure of $\mu_B$. Obviously this stays 
near constant within $0 < y < 1$, the heavy hadron radiation width as argued 
above. However, a different physics seems to set in, in each successive further 
unit rapidity interval. No use in widening the acceptance in a SHM analysis 
that seeks to establish the prevailing
($T$,$\mu_B$) point in the phase diagram, because $\mu_B$ increases, steeply, toward 
higher y where the valence quark density increases.

This is further demonstrate 
in the right panel of Fig.~\ref{fig:9} where the correlation between
K$^{-}/$K$^{+}$ and pbar/p is confronted with an interpolating SHM fit curve. The 
inferred $\mu_B$ shifts down from the $20$~MeV domain, corresponding to a 
relatively valence quark free midrapidity domain, to about $140$~MeV, smoothly 
interpolating, furtheron, to data points from the SPS domain.

To conclude about our above question of increasing the acceptance to secure the 
true Wroblewski ratio we see that from top SPS to top RHIC energies this is not 
helpful because each successive unit rapidity window reflects a differently 
composed hadronization source. This is the consequence of the hitherto little 
explored stopping mechanism that governs interpenetrating hadronic matter. At 
top LHC energies, the valence quark domain recedes far from midrapidity, and 
the acceptance is of no concern in this domain where $\mu_B$ is practically zero, 
and $T$ is constant (boost invariance). At low energies, on the other hand, the 
rapidity gap is so small that the energy and quark distributions fall mostly
within a single fireball radiation width: conditions yet to be explored
in detail.

\subsection{Core-Corona effects and canonical strangeness suppression}
\label{sec:6.2}

Glauber model calculations reveal that, even in central collisions, a small 
fraction of the target and projectile nucleons does not interact, at all, and a 
further about $10\%$ fraction of the participant nucleons interacts only once. 
Toward peripheral collisions that latter fraction increases toward about $25\%$. 
These single, minimum bias nucleon-nucleon collisions exhibit canonical 
strangeness suppression\cite{6,8}. The Wroblewski ratio of total strange to total 
nonstrange quarks in $p+p$ collisions is half that in central mass 200 $A+A$ 
collisions\cite{54}. In the SHM the transition from $p+p$ to $A+A$ occurs as a 
change-over from a canonical to a grand canonical ensemble that occurs, 
gradually, with increasing $N_{part}$\cite{6,8}. Singly strange/antistrange yields 
rise, by a factor of about two, relative to minimum bias $p+p$ at similar energy, 
and by factors of about 5 and 15, respectively, for multistrange 
hyperons\cite{54,55}. The single collision fraction of $A+A$ collisions stems from the 
diffuse surface regions of the nuclear Woods-Saxon nucleon density 
distributions. Thus one calls this the \textit{corona effect}\cite{6,54,56}. It reduces the 
strange to nonstrange hadron yield ratios in $A+A$ collisions, as compared to a 
global grand canonical equilibrium population. One attempt to account for this 
strangeness undersaturation is to introduce a strangeness undersaturation 
parameter $\gamma(s)^{S}$ into the partition functions of the GC Gibbs 
ensemble\cite{6}; $S = 1,2,3$ here. This is thus an additional fit parameter, in 
addition to the GC variables $V$, $T$ and $\mu_B$. It is found to amount to about 
$0.85$ to $0.90$ in central mass 200 collisions
at SPS and RHIC energies\cite{6}, increasing toward peripheral collisions\cite{57}.

We can illustrate a typical interplay between several such second order 
corrections to the standard GC approach. The effect of a narrow midrapidity
acceptance (see previous section) is expected to lead to an increase of the 
apparent Wroblewski ratio whereas the corona effect decreases it. The net 
effect might be a cancellation and, in fact, it was shown\cite{6} that the NA49 data 
for central $Pb+Pb$ collisions at top SPS energy (referring to a wider 
acceptance, of about $0 < y < 2$ required a $\gamma(s)=0.85$ whereas
$\gamma(s)=1$ as the SHM fit was restricted to a $0 < y < 1$ acceptance interval.

\subsection{The origin of canonical strangeness suppression}
\label{sec:6.3}

We address, next, a qualitatively different strangeness undersaturation effect, 
that governs elementary electron-positron annihilation to hadrons, and minimum 
bias $p+p$ collisions\cite{58,59}. We have shown in Fig.~\ref{fig:1} the canonical SHM fit to the 
LEP e$^{+}$--~e$^{-}$~annihilation data. It requires another, further strangeness 
suppression factor, $\gamma(s)=0.66$. Unfortunately the same term is used here as 
in the GC SHM analysis of $A+A$ collisions, confusing the issue as this 
suppression is of completely different origin. It can be understood from the 
Amati, Veneziano, Webber\cite{13,14} model of colour neutralization in elementary QCD 
hadronization that we referred to in Sec.~\ref{sec:4}. Here the primordial QCD partonic 
shower evolution is shown to end in a string-like configuration of successive 
complementary colour charges that locally neutralize into colour singlet 
clusters, with varying invariant masses. They decay, independently, into 
hadrons and resonances, according to their respective phase space weights. The 
clusters are distributed over the entire longitudinal phase space, and thus 
they are, in general, not causally connected\cite{58,59}. However, as shown by 
Becattini and Fries\cite{4}, the the Lorentz invariant single cluster multiplicities 
and net quantum numbers can be added into a single \textit{equivalent global cluster}, 
whose volume is the sum of the proper individual cluster volumes, turning out 
to be large enough to make the canonical ensemble a good approximation\cite{60}. 
However, the small volume , high microcanonical strangeness suppression 
(apparently due to the strange quark mass), prevailing in the individual 
cluster decays, gets transported into the equivalent global cluster strangeness 
content: a memory of the microscopic QCD dynamics leading to cluster formation. 
As the canonical SHM is ignorant of that microscopic pre-history it needs to be 
amended by an additional strangeness suppression parameter. It recalls that 
strangeness was produced in small, causally disconnected clusters, under high 
strangeness suppression.

From this, admittedly somewhat symbolic model, to hadron production from a QCD 
plasma, in $A+A$ collisions: how does a QGP hadronize? Clearly, the first stage 
of the above schematic model , local colour neutralization, has to be in 
common. We are not aware of any theoretical argument as to why this should not 
be a short-range, local process, imbedded into the overall collective 
expansion/cooling dynamics as the plasma approaches the critical or 
pseudocritical energy density domain. Furthermore it is plausible that the 
cluster density in phase space should now be about A times higher. So the 
phenomenon of isolated, single cluster hadronization that we addressed above 
under the symbolic term \textit{causally not connected}
should disappear, leading to a quantum number coherent large equivalent 
super-cluster decay, that is mirrored in the success of the Grand Canonical, 
large volume SHM description of the $A+A$ hadron output. As we have seen from the 
BRAHMS results in Fig.~\ref{fig:9}, such fictitious superclusters can not extend over all 
longitudinal phase space, as successive intervals in rapidity space reveal a GC 
hadronization under different ($T$,$\mu_B$) conditions. But, on the other hand, the 
genuine small cluster volume strangeness suppression, known from elementary 
(minimum bias) collisions, has disappeared here, hinting at a much larger, 
coherent decay system, in accord with the above schematic super-cluster 
picture.

In all, the reader will notice, however, that we employ here a highly symbolic 
terminology. Clusters, and super-clusters, are not yet known from genuine QCD 
calculations. They may be understood as an extrapolation from the concept of 
massive hadronic resonances. However, the formal three-step idea may reflect, 
in a minimal way, the three steps expected from fundamental QCD: colour 
neutralization corresponding to the confinement transition, hadronic (cluster) 
mass generation symbolizing QCD chiral symmetry breaking, and the final quantum 
mechanical decay into on-shell hadrons/resonances representing the final 
tunneling effect, with its intrinsic phase space dominance, a la Fermis Golden 
Rule, that makes the apparent hadro-chemical equilibrium plausible.

\subsection{Onset of Grand Canonical hadron production at the LHC}
\label{sec:6.4}

It has been predicted, before the startup of the LHC, that at $\sqrt(s) = 7~$TeV 
the midrapidity production of hadrons in $p + p$ collisions might approach the 
grand canonical limit\cite{60} of the SHM description. These expectations can now be 
checked with ALICE $p + p$ data\cite{61} on strangeness production vs. multiplicity 
density $dN(ch)/d\eta$. In the cluster-terminology employed in the previous 
section the advent of GC strangeness saturation would have to imply an 
increasing overlapp (or \textit{causal connection}) among the microscopic singlet 
clusters, in longitudinal phase space. Looking at the midrapidity densities of 
charged hadrons we find $dN(ch)/d\eta = 3.3$ in LEP electron-positron 
annihilation (Fig.~\ref{fig:1}), and $6.0$ in minimum bias $p + p$ at $7$~TeV. The latter 
value\cite{61} is quite in line with the $p + p$ systematics at lower energies\cite{54}. As 
the former case did require the application of the strangeness suppression 
factor $0.66$, characteristic of elementary collisions at such lower 
energies\cite{60}, one would not expect a substantial, qualitative change in $p + p$ 
at the LHC as the cluster density increases by less than a factor of two. 
\begin{figure}[t]
%\sidecaption
\centering
\includegraphics[width=.4\textwidth]{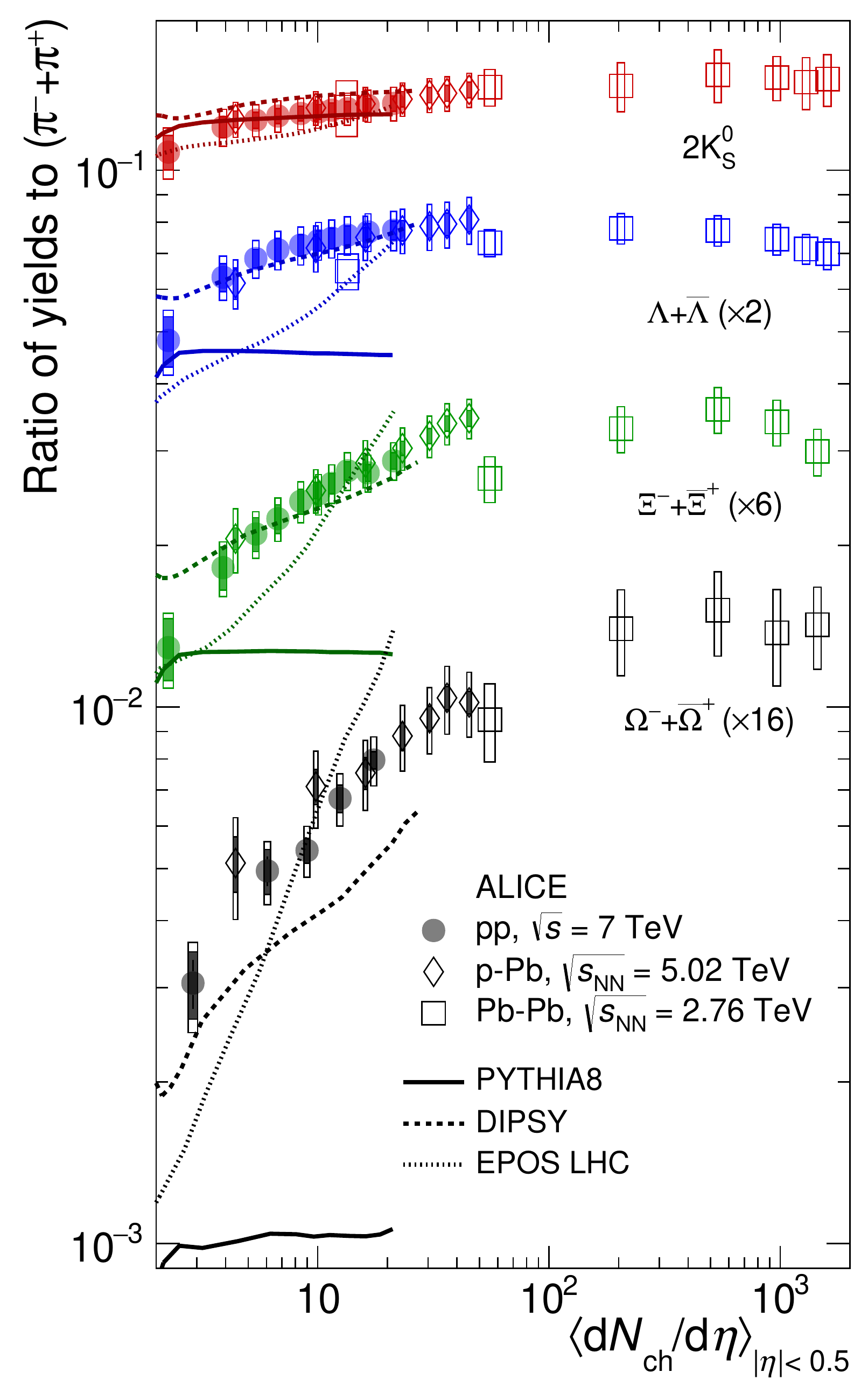}
%\picplace{5cm}{2cm} % Give the correct figure height and width in cm
%
\caption{ALICE results for $p+p$ collisions at LHC energy, $\sqrt{s}=7$~TeV\cite{62},
showing the multiplicity dependence of strange hadron to pion yields, matching 
to results from $p + Pb$ and $Pb + Pb$ collisions.}
\label{fig:10}       % Give a unique label
\end{figure}
However, the ALICE data shown in Fig.~\ref{fig:10} imply that, on the contrary, 
significant strangeness enhancement (the reversal of strangeness suppression) 
occurs, with increasing midrapidity charged hadron density. Of course in the 
highest multiplicity bin, analysed here, we have $dN_{ch}/d\eta = 21.3$, now in 
fact 6.5 times higher than in e$^{+}$-e$^{-}$ annihilation at LEP. Also the 
enhancement effect increases strongly with ascending strangeness, quite 
resembling the strangeness hierarchical increase in the course of a transition 
from the canonical to the grand canonical ensemble\cite{8,54}. However, we also see 
in Fig.~\ref{fig:10} that the Omega to pi ratio, for example, grows by a further factor of 
about two, toward semi-central $Pb + Pb$ collisions at $2.76$~TeV. As the latter 
reaction should represent full GC strangeness saturation, we finally conclude 
that, even in the high multiplicity density windows of $p + p$ at LHC, we see no 
fully developed GC behavior but the onset of it. A GC SHM fit to these data 
would thus feature a non-unity, genuine $\gamma(s)$ factor.

A final remark concerning $p+p$ data as a reference for hadron production in $A+A$ collisions, as expressed in the nuclear modification factor $R_{AA}$. Note 
that this is commonly defined as the ratio between $A+A$ cross section and the 
corresponding minimum bias $p+p$ cross section scaled with the number of binary 
collisions, corresponding to the considered $A+ A$ centrality class. However, 
this ratio is only applicable at high transverse momentum, well above the 
domain of thermal bulk production. Here the hadron multiplicities scale with 
the volume of the emitting source, proportional to $N_{part}$, and not with the 
Glauber binary collision number that increases like $N_{part}^{4/3}$. At high 
$p_t$ there should be no problem with employing $R_{AA}$.

\section{Conclusions}
\label{sec:7}

The fundamental interest in Statistical Model analysis stems from the 
hypothesis that the derived freeze-out points in the ($T$,$\mu_B$) phase diagram 
should, at modest $\mu_B$, coincide with the QCD hadronization points, thus 
revealing the parton-hadron transformation line of QCD: one of the principal 
characteristics of thermal QCD, in the non-perturbative domain. It was the 
purpose of this article to demonstrate that SHM studies are converging toward 
this goal. One needs, however, to take care of numerous second order concerns 
in the application of the grand canonical SHM. From among these, we have 
addressed final state annihilation corrections, experimental effects stemming 
from  lack of feed-down corrections arising from unresolved, secondary vertex 
weak hyperon and
antihyperon decays, and proper handling of acceptance and core-corona effects. 
Further problems require attention, such as finite hadron volume 
corrections\cite{62} and the clarification of the relationship between the SHM and 
coalescence model approaches\cite{4,63}. Finally the validity of the single 
freeze-out model\cite{8,47} still awaits a clarification.
Turning to the idea\cite{17} to replace, or substitute the traditional SHM analysis 
by considering the overlap between Lattice QCD and Hadron-Resonance-Gas(HRG) 
model calculations: this should provide for a alternative method to locate the 
\mbox{(pseudo-)critical} temperature domain over which partonic degrees of freedom 
convert to the hadronic ones\cite{18}. One can also analyze, directly, the first 
data concerning higher order grand canonical fluctuations of conserved quantum 
numbers, such as skewness and kurtosis, in terms of Lattice predictions for 
higher order susceptibilities\cite{19,20,33,64}. We have argued that not all of the 
the various quantities, addressed in this framework (to which we could only 
provide a passing glance), should necessarily reveal a common freeze-out 
temperature\cite{65}, given the broad band of temperatures encountered in a 
cross-over phase transformation as is encountered at modest $\mu_B$. First order 
quantities, like energy and entropy densities (closely related to the hadron 
number densities considered in the SHM), might freeze out sooner than fourth 
order susceptibilities. Finally the Hadron-Resonance Gas approximation to the 
latter type of quantities may lack validity\cite{62}.
\begin{figure}[t]
%\sidecaption
\centering
\includegraphics[width=.85\textwidth]{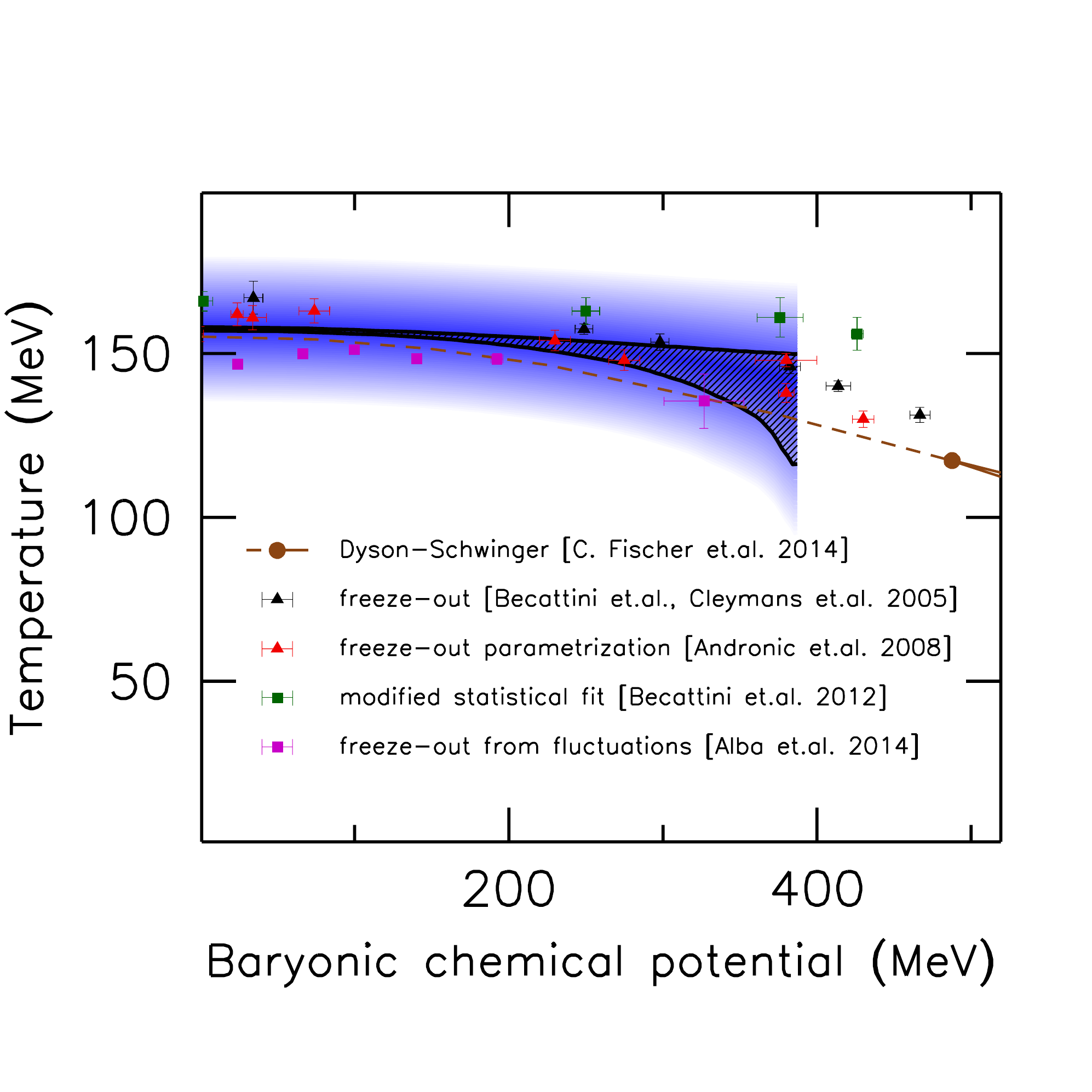}
%\picplace{5cm}{2cm} % Give the correct figure height and width in cm
%
\caption{The QCD phase diagram at finite baryochemical potential showing the 
domain of the crossover transition, confronted with various entries from
SHM and Dyson-Schwinger analysis(see Ref.\cite{66} for detail).}
\label{fig:11}       % Give a unique label
\end{figure}
To wrap up on this new research field, of HRG-Lattice QCD-data analysis, and 
also the recent attempts in Lattice QCD to pin down the pseudo-critical line 
(or domain) in the QCD phase diagram, we show in Fig.~\ref{fig:11} the result of a recent 
study of the phase diagram, by Bellwied \textit{et al.}\cite{66}. The relatively broad domain 
of the cross-over transition in the ($T$,$\mu_B$) plane, reaching up to limit of 
validity of the employed extrapolation method for finite $\mu_B$ domains, at 
about $400$~MeV, is shown to envelope individual entries stemming from SHM and 
other approaches(see Ref.\cite{66}). We conclude that a detailed description of the 
QCD parton-hadron transition, far into the finite $\mu_B$ domain, appears within 
reach now, a formidable task commensurate to its fundamental importance.
What remains, relatively much less explored, is the high $\mu_B$ domain.
A rather abrupt turn-over occurs\cite{7} at a $\mu_B$ of about $400$~MeV, but the 
scarcity of data in this domain prevents us from answering several interesting 
questions concerning the high $\mu_B$ domain:

\begin{enumerate}
\item If one considers the lattice extrapolations to apply, at all, to such high 
$\mu_B$ we might ascribe the the steepening slope to higher than quadratic terms 
in the lattice Taylor expansion.

\item If the enigmatic critical point of QCD exists, at all, it should occur\cite{16} 
at high $\mu_B$, and influence the sequence, and position of the hadronization 
points (their dependence on the incident energy of the collision). This might 
occur, either, due to the onset, at the critical point, of a first order phase 
transition, or to the hypothetical focusing effect on the systems expansion 
trajectories\cite{17}.

\item A further untested suggestion would explain the rather steep fall-off toward 
AGS energy. If the so-called quarkyonic matter domain of QCD\cite{19} exists, and 
sets in at the temperature of the critical point, one would expect a second QCD 
phase boundary line, to turn down, steeply, from the continuing deconfinement 
line, and become the site of hadronization.

\item Finally, to the more trivial side, we might expect to turn away from the QCD 
transition line because the collisional volume does not (or not predominantly) 
enter the deconfined phase, to begin with. However, recent hadron transport 
model studies\cite{67} consistently predict maximum energy densities well above
one GeV/fm$^{3}$ to be reached at this energy, indicating an onset of 
deconfinement\cite{18} to occur at lower energies.
\end{enumerate}
We conclude that, neither the theoretical foundation, nor the approaches toward 
an experimental evidence, have been receiving even remotely such an attention 
recently as we witnessed in the attempts to work out the QCD phase diagram at 
modest $\mu_B$, where Lattice QCD is applicable.

\bibliographystyle{ws-rv-van}
%\bibliography{ws-rv-sample}
\input{referenc}

%\printindex[aindx]                 % to print author index
%\printindex                         % to print subject index
\end{document}

%% file: referenc.tex
%%%%%%%%%%%%%%%%%%%%%%%% referenc.tex %%%%%%%%%%%%%%%%%%%%%%%%%%%%%%
% sample references
% %
% Use this file as a template for your own input.
%
%%%%%%%%%%%%%%%%%%%%%%%% Springer-Verlag %%%%%%%%%%%%%%%%%%%%%%%%%%
%
% BibTeX users please use
% \bibliographystyle{}
% \bibliography{}
%